\documentclass[cameraready]{Interspeech}

\newcommand{\vtag}{v_{\text{tag}}}
\newcommand{\vnull}{v_{\text{cf}}}
\newcommand{\vunc}{v_{\text{unc}}}
\newcommand{\vcfg}{v_{\text{cfg}}}

\title{Where More Than Which: Distilling Tag-Contrastive Guidance for Non-Verbal Vocalizations}

\author[affiliation={1}]{Guoqiang}{Hu}
\author[affiliation={1,2}]{Yizhou}{Peng}
\author[affiliation={1,2}]{Changsong}{Liu}
\author[affiliation={1}]{Yi-Wen}{Chao}
\author[affiliation={1}]{Yuxi}{Wang}
\author[affiliation={1}]{Yihao}{Wu}
\author[affiliation={1}]{Mengjun\nobreak}{Wu}
\author[affiliation={1}]{Qianen}{Zhang}
\author[affiliation={3}]{Chenyang}{Lyu}
\author[affiliation={1}]{Eng Siong}{Chng}
\address{$^1$Nanyang Technological University, Singapore\\
$^2$Alibaba-NTU Global e-Sustainability CorpLab, Nanyang Technological University, Singapore\\
$^3$Alibaba Group, China}
\email{guoqiang002@e.ntu.edu.sg}


\begin{document}
\maketitle

\begin{abstract}
An inline vocalization tag specifies both where a sound should occur and which sound it should be. In VoxCPM2, adding a tag brings non-verbal vocalization (NV) placement close to that in real recordings, yet the requested class occurs about 20 percentage points less often. Likelihood fine-tuning strengthens the model's preference for the correct tag under teacher forcing but yields little improvement in sampled class realization. We introduce tag-contrastive guidance (TCG), which compares predictions for two texts differing only at the tag. Distilling the contrast into the flow-matching target lets the student use ordinary decoding. The distilled model raises class realization from 40.6\% to 50.5\% under an NV-aware recognizer and from 36.5\% to 52.1\% under an audio LLM judge. Character and word error rates rise by 0.9 and 0.3 points. In blind comparisons of NV accuracy, ten listeners preferred the distilled model in 48\% of judgments and the baseline in 15\%. Audio samples are available at \mbox{\href{https://coding-sharks.github.io/TCG-VoxCPM2-demo/}{coding-sharks.github.io/TCG-VoxCPM2-demo}}.
\end{abstract}

\section{Introduction}
Laughter, sighs, coughs and breaths convey affect and help speakers manage interactions~\cite{nvspeech,nonverbaltts}. In expressive text-to-speech (TTS), users can request these sounds with inline tags such as [laugh]. Recent corpora~\cite{nvspeech,nvs38k,nonverbaltts} and benchmarks~\cite{nvvsuperbench,nvbench} support evaluation of tag control. In the NVVSpeech Challenge 2026~\cite{nvvchallenge}, however, no system scored above 3 out of 5 on listener-rated NV accuracy.

An overall NV accuracy score leaves two questions open. Does the system produce a vocalization at the tag (\emph{placement})? Does it produce the requested class (\emph{class realization})? A sigh at a position marked [laugh] can preserve the timing of an utterance while changing how a listener interprets the words that follow. We synthesize each test sentence with and without its tag using the same seed and compare these pairs with real recordings. The untagged output, or \emph{twin}, shows how much control the tag adds. For the challenge baseline, adding the tag closes most of the placement gap, while class realization remains about 20 points below real recordings.

Existing approaches improve NV training data~\cite{nvspeech,nvs38k,synparaspeech}, rebalance classes~\cite{tencenthy} or use evaluator feedback for sample selection and preference optimization~\cite{flowdpo,nvdpo}. A training recording supplies evidence for both decisions. Predicting how a laugh continues once it is audible, however, makes a different demand on the model from initiating a laugh in response to text. Our eleven likelihood fine-tuning variants yield at most modest gains. Under teacher forcing, a stronger preference for the true tag does not reliably predict better class realization during synthesis (Secs.~\ref{sec:exp} and~\ref{sec:tf}).

We use the tag itself to define a \emph{counterfactual}. Holding the words and audio history fixed, we compare the model's prediction for the requested tag with its prediction under a different tag or a deletion. This difference supplies tag-contrastive guidance (TCG) at inference. We also distil it into the flow-matching target and compare this with training on guided samples or adapting condition contrastive alignment (CCA)~\cite{cca}.

Our contributions are threefold. First, paired synthesis and real recordings separate placement from class realization in VoxCPM2~\cite{voxcpm2}. Second, we use a tag-local contrast both to guide inference and to train a student that retains standard decoding. Finally, we show that a stronger preference for the true tag under teacher forcing does not reliably predict better NV class realization during synthesis.

We release the TCG-VoxCPM2 weights\footnote{\urlstyle{same}\Urlmuskip=0mu\relax\url{https://huggingface.co/airguoqiang/TCG-VoxCPM2}} and source code\footnote{\urlstyle{same}\Urlmuskip=0mu\relax\url{https://github.com/coding-sharks/TCG-VoxCPM2}}.

\section{Related work}
\label{sec:related}
\textbf{NV synthesis.} NV-annotated corpora~\cite{nvspeech,nvs38k,nonverbaltts}, spliced NV clips~\cite{synparaspeech} and dedicated tokens~\cite{tencenthy} provide supervision for tag control. Benchmarks~\cite{nvvsuperbench,nvbench} and quality predictors such as NVMOS~\cite{nvmos} support evaluation. We use tagged and untagged pairs to isolate the effects on placement and class identity.

\textbf{Guidance.} Classifier-free guidance (CFG)~\cite{cfg} adapts classifier guidance~\cite{classifierguidance} to conditional generation and is widely used in flow-matching TTS~\cite{voicebox}. Guidance can isolate individual conditions~\cite{selectivecfg,ccgcfg,eccd}, reshape a guidance vector~\cite{apg} or use a weaker model as a reference~\cite{autoguidance}. Our comparison draws on contrastive prompts in diffusion~\cite{composable,perpneg,contrastiveguidance,contrastivecfg} and context-aware decoding~\cite{cad}. Holding the surrounding words fixed lets us compare competing NV classes in the same sentence.

\textbf{Distillation and contrastive objectives.} Prior work distils guided predictions~\cite{guiddistill,modelguidance,cfgfreetts} or changes the loss through condition contrasts~\cite{cca,gft}, sample contrasts~\cite{deltafm,robustspeechflow} or likelihood discrimination~\cite{ddo}. We adapt model guidance~\cite{modelguidance} by using a tag-local counterfactual in the teacher target. Comparison with CCA tests whether placing the contrast in the target or the loss matters for NV realization.

\section{Task, evaluation and diagnosis}
\label{sec:diag}
\textbf{Task and metrics.} Each inline tag requests one of 16 NV classes. Placement measures whether any NV occurs at the tag position. Class realization counts an output as a success if an evaluator detects the requested class anywhere in it. We also measure \emph{at-tag} realization, which requires the recognizer to detect that class within two English words or two Mandarin characters of the tag, and \emph{sole-NV} realization, which requires the requested class to be the only NV heard. The NVVSpeech Track-2 test set contains 800 Mandarin and 800 English sentences, with about 100 tags per class. The official baseline is VoxCPM2 (2B parameters, CFG scale $w{=}2$), fine-tuned by the organizers for ten epochs on NV-tagged speech~\cite{nvvchallenge,voxcpm2}.

\textbf{Evaluators.} We use two complementary measures of NV realization~\cite{flowdpo,nvdpo,tencenthy,nvbench}. The Track-1 Whisper-medium \emph{recognizer} places NV tags in its transcript, allowing us to score both class and position. The Qwen3-Omni-30B-A3B-Instruct \emph{judge}~\cite{qwen3omni} listens without the input text and lists the NVs it hears. On 350 natural NV recordings, the recognizer and judge detected 65\% and 49\% of real NVs, respectively. On 400 NV-free recordings, the judge reported an NV in 4\%, while the recognizer inserted tags in most NV-free Mandarin recordings. The judge also over-reported sighs. We therefore compare systems within each evaluator, using untagged twins and real recordings as reference points. Agreement across evaluators tests whether gains extend beyond transcript-based scoring.

We measure lexical accuracy with Paraformer character error rate (CER) for Mandarin~\cite{paraformer} and Whisper-large-v3 word error rate (WER) for English~\cite{whisper}. PCER is the edit distance between requested and recognized NV-tag sequences per requested tag~\cite{nvdpo}. DNSMOS P.835 OVRL estimates speech quality~\cite{dnsmos}.

\begin{table}[t]
\centering
\caption{NV detections (\%) on 1,366 single-tag test sentences and 305 recordings of the same classes, excluding breath and sigh. No tag denotes the untagged twins. Req. denotes the requested class, None denotes no NV, and Sigh and Gasp are substitutions. The recognizer scores within two English words or two Mandarin characters of the tag. Italics mark reference rates and bold marks the better system.}
\label{tab:diag}
\footnotesize
\setlength{\tabcolsep}{2.6pt}
\begin{tabular}{@{}lcccc@{\hspace{9pt}}cccc@{}}
\toprule
 & \multicolumn{4}{c}{Recognizer, at the tag} & \multicolumn{4}{c}{Judge, anywhere} \\
\cmidrule(lr){2-5}\cmidrule(l){6-9}
 & Req.\,$\uparrow$ & None & Sigh & Gasp & Req.\,$\uparrow$ & None & Sigh & Gasp \\
\midrule
\textit{Recordings} & \textit{54.8} & \textit{25.2} & \textit{1.3} & \textit{3.6} & \textit{51.1} & \textit{4.9} & \textit{18.7} & \textit{4.3} \\
\textit{No tag} & \textit{6.3} & \textit{58.0} & \textit{0.0} & \textit{5.1} & \textit{17.3} & \textit{47.4} & \textit{20.3} & \textit{3.7} \\
\midrule
Baseline & 34.7 & 28.3 & 0.7 & 14.9 & 32.4 & 11.6 & 38.8 & 7.1 \\
Distilled & \textbf{46.9} & \textbf{26.1} & \textbf{0.0} & \textbf{3.8} & \textbf{48.9} & \textbf{11.3} & \textbf{26.9} & \textbf{1.8} \\
\bottomrule
\end{tabular}
\end{table}

\textbf{Where more than which.} On the single-tag examples in Table~\ref{tab:diag}, adding the tag reduces the share of positions with no recognized NV from 58\% to 28\%, close to the 25\% in real recordings. Sound-filled gaps of at least 0.15\,s occur at 50.8\% of 770 tagged positions versus 20.5\% in untagged twins. Yet requested-class detections remain about 20 points below real recordings under both evaluators. On the full test set, the recognizer counts 54 substitutions and 5 omissions per 100 requested tags. Class confusion accounts for most of these errors.

\section{Tag-contrastive guidance}
\label{sec:tcg}
\begin{table*}[t]
\centering
\caption{Class realization (\%) on 1,600 test sentences and 1,200 context-free items. Rec. denotes the recognizer. Sentence-paired bootstrap 95\% CIs for test-set differences from baseline are 4.0--4.9 points wide and exclude zero except for global CFG (Rec.), the likelihood control (Judge) and the best likelihood variant (Rec.); distilled CFG (Rec.) is borderline. Bold and underline mark the best and second-best values. $^\dagger$Strength selected on NVV-SuperBench development data~\cite{nvvsuperbench}.}
\label{tab:res}
\footnotesize
\setlength{\tabcolsep}{5pt}
\begin{tabular}{lcccccccc}
\toprule
 & \multicolumn{2}{c}{Test set (\%) $\uparrow$} & \multicolumn{2}{c}{Context-free (\%) $\uparrow$} & \multicolumn{3}{c}{Error (\%) $\downarrow$} & Quality $\uparrow$ \\
\cmidrule(lr){2-3}\cmidrule(lr){4-5}\cmidrule(lr){6-8}\cmidrule(l){9-9}
System & Rec. & Judge & Rec. & Judge & CER & WER & PCER & DNSMOS \\
\midrule
Baseline ($w{=}2$) & 40.6 & 36.5 & 47.3 & 20.5 & 6.15 & 3.09 & 84.7 & \underline{3.04} \\
\midrule
\multicolumn{9}{l}{\textit{Guidance at inference}} \\
Global CFG ($w{=}4$) & 39.8 & 40.5 & 48.5 & 23.2 & \underline{5.55} & \underline{2.93} & 81.6 & 2.86 \\
TCG ($\lambda{=}8^\dagger$) & \textbf{50.9} & \underline{51.6} & \textbf{66.4} & \underline{37.2} & 7.75 & 3.91 & 74.2 & 2.99 \\
\midrule
\multicolumn{9}{l}{\textit{Likelihood training ($\lambda'{=}0$)}} \\
Likelihood control (400 steps) & 37.8 & 37.1 & 40.2 & 17.9 & 5.80 & 2.95 & 86.9 & \underline{3.04} \\
Best of eleven (NV-weighted loss, 1,200 steps) & 41.5 & 41.3 & 42.9 & 21.7 & \textbf{5.45} & \textbf{2.88} & 82.8 & 3.03 \\
\midrule
\multicolumn{9}{l}{\textit{The contrast in training, by route}} \\
Data: own TCG-guided samples & 46.1 & 47.7 & 61.6 & 33.5 & 6.43 & 3.82 & \underline{73.9} & 2.94 \\
Loss: CCA$^\dagger$ & 48.0 & 44.6 & 57.2 & 30.3 & 6.14 & 3.12 & 81.4 & 3.03 \\
Target: distilled CFG ($\lambda'{=}2$) & 42.9 & 41.8 & 49.2 & 24.6 & 5.78 & 3.29 & 74.6 & 2.84 \\
Target: distilled student ($\lambda'{=}4^\dagger$) & \underline{50.5} & \textbf{52.1} & \underline{64.9} & \textbf{37.9} & 7.06 & 3.37 & \textbf{72.4} & \textbf{3.09} \\
\bottomrule
\end{tabular}
\end{table*}

\textbf{Guidance at inference.} Let $x$ denote the tagged text and $\tilde x$ the same text with each NV tag replaced by a counterfactual token $c$. VoxCPM2 generates speech autoregressively in 160\,ms patches. For a given noisy patch $y_t$, we evaluate its flow-matching head under both texts with the same audio history. Each velocity describes an update to the same patch at the same flow time. TCG adds a scaled tag contrast to the CFG velocity,
\begin{equation}
v = \vcfg + \lambda\,P(\vtag-\vnull),\qquad \vcfg=\vunc + w(\vtag-\vunc),
\label{eq:tcg}
\end{equation}
where $\vtag=v(y_t\mid x)$, $\vnull=v(y_t\mid\tilde x)$ and $\vunc$ is the unconditional prediction. In Eq.~\ref{eq:tcg}, the ordinary CFG difference compares the complete tagged sentence with an unconditional prediction. Increasing $w$ amplifies every change induced by the text, including those associated with the surrounding words. The second difference asks what changes when the words stay fixed and the requested NV changes. For example, both the [laugh] and [sigh] versions of a sentence request a vocalization at the same position. The subtraction cancels updates common to both predictions and retains those that depend on the requested sound. The edit is local to the tag in the text, although its acoustic effects can extend into adjacent speech.

$P$ is either the identity or a projection that removes the component of the contrast parallel to the guided velocity $\vcfg$, following APG~\cite{apg}. With the identity, the contrast can also rescale the guided velocity. With projection, it adds only a component orthogonal to $\vcfg$, with strength set by $\lambda$. For exact velocities, the unprojected contrast is $-t/(1-t)$ times the log-odds gradient $\nabla_{y_t}\log[p_t(x\mid y_t)/p_t(\tilde x\mid y_t)]$, with $t{=}1$ at noise, so sampling favors the requested tag over the counterfactual.

\textbf{Counterfactuals.} We either delete the tag (replacing it with a comma), substitute a different NV class chosen at random, or apply a \emph{fixed fallback} rule. The rule replaces [breath] with [sigh] and every other tag with [breath]. Substitution holds the request for an NV at that position fixed. A random class changes the competitor, whereas the fixed rule gives each tag a consistent alternative. For [laugh], for example, the fixed comparison asks which update favors a laugh over a breath. With deletion, the competing text has only punctuation at that position. Its guidance can therefore reflect the presence of an NV as well as its identity.

\textbf{Distilling the contrast.} To avoid evaluating the counterfactual branch at inference, we distil the tag contrast into the training target. With $y_0$ a clean training patch, $z$ Gaussian noise and $y_t=(1-t)y_0+tz$ their mixture at flow time $t$, the ordinary flow-matching target $z-y_0$ becomes
\begin{equation}
u = (z-y_0) + \lambda'\big(v_{\mathrm{T}}(y_t\mid x)-v_{\mathrm{T}}(y_t\mid\tilde x)\big),
\label{eq:distill}
\end{equation}
where $v_{\mathrm{T}}$ is the frozen initial model. The first term gives the training velocity for the observed patch and its sampled noise. For that same patch, noise and flow time, we evaluate the teacher twice, changing only the tag condition. If the teacher predicts the same velocity under both texts, the example keeps its ordinary target. Where the predictions differ, the second term shifts the target by an amount that depends on how the teacher responds to the edit. This gives the student a patch-level training signal without requiring a recording of the counterfactual version of the sentence.

The student predicts this modified target from the requested text. Training examples with dropped text retain the ordinary target, so the tag contrast enters through the conditional prediction. At inference, the student uses \emph{plain decoding}, ordinary CFG at $w{=}2$ with no counterfactual branch. CFG scales the learned contrast by $w$, giving an effective unprojected strength of approximately $w\lambda'$. Substituting the unconditional teacher prediction for the counterfactual gives distilled CFG. Setting $\lambda'{=}0$ recovers the ordinary target (the likelihood control). Together these controls distinguish the effect of adding guidance to the target from the effect of choosing a tag-local contrast.

\textbf{Contrastive loss and guided samples.} CCA~\cite{cca} favors matched over mismatched conditions relative to a frozen reference. We adapt it to flow matching following Diffusion-DPO~\cite{diffdpo}, replacing the requested tag with a randomly chosen different class. For the guided-sample comparison, we train on the baseline's TCG-guided outputs using the ordinary likelihood objective. Here the student observes the completed guided utterance. Target distillation instead uses the difference between the teacher's predictions at the noisy training patch.

\section{Experiments}
\label{sec:exp}
\textbf{Setup.} All students start from the Track-2 baseline. Repeating rare classes yields 65.9k examples from 46.6k recordings across seven public NV corpora (including those in~\cite{nonverbaltts,nvs38k,synparaspeech}). Defaults are 400 steps, batch size 126 and learning rate $5{\times}10^{-6}$. TCG uses the fixed fallback, projection and $\lambda{=}8$. Distillation uses the same fallback and $\lambda'{=}4$.

\textbf{Likelihood fine-tuning.} Eleven variants cover longer training (up to 3,000 steps), learning rate, class balancing, LoRA, reference clips, NV-weighted loss and self-training. None improves realization by more than 5 points under either evaluator (Table~\ref{tab:res}).

\textbf{Tag-local guidance and distillation.} With standard CFG decoding ($w{=}2$), the distilled student raises recognizer realization from 40.6\% to 50.5\% and judge realization from 36.5\% to 52.1\% (Table~\ref{tab:res}). PCER falls from 84.7\% to 72.4\%, and DNSMOS remains comparable (3.04 and 3.09). Mandarin CER rises from 6.15\% to 7.06\% and English WER from 3.09\% to 3.37\%. Inference-time TCG gains 10.3 points under the recognizer and 15.1 under the judge. Distillation achieves comparable realization without the extra counterfactual branch, with smaller increases in CER and WER (0.9 and 0.3 points versus 1.6 and 0.8 for TCG). Global CFG and distilled CFG gain at most 5.3 points under either evaluator. The source of the contrast thus matters both when it is applied during sampling and when it is transferred to a student.

On the single-tag subset (Table~\ref{tab:diag}), the share of positions with no recognized NV changes little between baseline and distilled outputs (28.3\% and 26.1\%), while the share with the requested class rises from 34.7\% to 46.9\%. The judge likewise reports fewer substitutions by sighs and gasps. Among the three tag-local counterfactuals, realization gains are within 4 points of the fixed fallback. The planned comparisons (inference and $\lambda'{=}2$) show no significant difference (Holm-adjusted McNemar tests). Deletion incurs the largest lexical cost, up to 2.0 points in CER and 2.7 in WER.

\textbf{Realization without contextual cues.} TCG and the distilled student retain gains of 9 to 15 points under the stricter sole-NV criterion. We also tested 1,200 \emph{context-free} items comprising test sentences retagged with an unrelated class and neutral carrier sentences with a common tag position. In an ordinary sentence, the words alone may suggest a laugh or a sigh. Swapping the tag breaks that association, while a neutral carrier lets different classes occupy exactly the same lexical context. A gain on these items gives more direct evidence that the model follows the explicit request. The judge's baseline realization falls from 31.9\% on original sentences of the same 12 classes to 20.5\% on these items. Both TCG and the distilled student gain at least 16.7 points over the context-free baseline under each evaluator (Table~\ref{tab:res}).

\textbf{Other systems and NV-Bench.} We evaluated six open-source systems with their default decoding settings and voices~\cite{stepeditx,cosyvoice3,bark,maya1,orpheus,drama3}. Five support 4 to 7 of the 16 classes. Drama~3 accepts free-form tags. On each system's supported subset, both evaluators score the distilled student above the likelihood control (Table~\ref{tab:sys}). Only Maya1 exceeds the student, under the recognizer. On the full NVVSpeech test set, Drama~3 scores 49.5\% under the judge versus 52.1\% for the student, a nonsignificant difference. NV-Bench~\cite{nvbench} provides 529 sentences with human recordings across four shared classes. The distilled student improves on baseline by 5.5 points under the recognizer and 4.2 under the judge (both CIs exclude zero), reaching rates comparable to the recordings. For TCG the changes are $+3.2$ and $-1.3$ points, neither significant. Of the 279 English sentences, 8 have WER above 60\% with TCG, compared with 2 for the baseline and none for the student.

\begin{table}[t]
\centering
\caption{Class realization (\%) on each system's supported subset. Sys. denotes the external system, Base the VoxCPM2 baseline, Lik. the likelihood control and Ours the distilled student. Cl. gives the number of supported classes and $^{\mathrm{en}}$ indicates English only. Bold and underline mark the top two systems within each row and evaluator, excluding recordings.}
\label{tab:sys}
\footnotesize
\setlength{\tabcolsep}{1.05pt}
\begin{tabular}{@{}lcc@{\hspace{4pt}}cccc@{\hspace{4pt}}cccc@{}}
\toprule
 & & & \multicolumn{4}{c}{Recognizer} & \multicolumn{4}{c}{Judge} \\
\cmidrule(lr){4-7}\cmidrule(lr){8-11}
System & Cl. & $n$ & Sys. & Base & Lik. & Ours & Sys. & Base & Lik. & Ours \\
\midrule
\multicolumn{11}{l}{\textit{NVVSpeech test set}} \\
Step-Audio-EditX & 5 & 500 & 29.4 & \underline{29.8} & 21.9 & \textbf{39.1} & \underline{56.6} & 51.5 & 50.8 & \textbf{69.4} \\
CosyVoice3 & 6 & 600 & 25.6 & \underline{43.5} & 37.4 & \textbf{49.8} & 33.1 & \underline{43.6} & 43.5 & \textbf{57.3} \\
Bark & 4 & 400 & 17.2 & \underline{35.9} & 32.0 & \textbf{38.3} & 61.2 & \underline{65.3} & 62.9 & \textbf{77.4} \\
Maya1$^{\mathrm{en}}$ & 5 & 250 & \textbf{28.1} & 17.0 & 17.4 & \underline{22.5} & 39.9 & 49.8 & \underline{52.6} & \textbf{64.4} \\
Orpheus-3B$^{\mathrm{en}}$ & 7 & 350 & 26.9 & \underline{29.7} & \underline{29.7} & \textbf{37.4} & 29.7 & 40.5 & \underline{43.6} & \textbf{53.5} \\
Drama 3 & 16 & 1600 & 39.0 & \underline{40.6} & 37.8 & \textbf{50.5} & \underline{49.5} & 36.5 & 37.1 & \textbf{52.1} \\
\midrule
\multicolumn{11}{l}{\textit{NV-Bench, 4 classes}} \\
Recordings & 4 & 529 & \textit{44.0} & \underline{42.3} & 39.2 & \textbf{47.8} & \textit{53.1} & \underline{51.1} & 48.9 & \textbf{55.3} \\
Step-Audio-EditX & 4 & 529 & 37.8 & \underline{42.3} & 39.2 & \textbf{47.8} & 42.7 & \underline{51.1} & 48.9 & \textbf{55.3} \\
CosyVoice3 & 4 & 529 & 25.7 & \underline{42.3} & 39.2 & \textbf{47.8} & 24.4 & \underline{51.1} & 48.9 & \textbf{55.3} \\
Bark & 2 & 229 & 13.8 & \underline{53.5} & 49.6 & \textbf{60.4} & 44.6 & \underline{85.4} & 80.4 & \textbf{95.4} \\
Maya1$^{\mathrm{en}}$ & 3 & 198 & \textbf{23.7} & 18.7 & 13.6 & \underline{22.2} & 54.1 & \underline{54.9} & 47.9 & \textbf{62.6} \\
Orpheus-3B$^{\mathrm{en}}$ & 3 & 172 & 9.4 & \underline{25.7} & 23.3 & \textbf{35.6} & 53.5 & \underline{62.4} & 61.4 & \textbf{74.8} \\
Drama 3 & 4 & 529 & 32.6 & \underline{42.3} & 39.2 & \textbf{47.8} & 46.6 & \underline{51.1} & 48.9 & \textbf{55.3} \\
\bottomrule
\end{tabular}
\end{table}

\textbf{Where to apply the contrast during training.} Guided-sample training gains 5.5 points under the recognizer and 11.2 under the judge. CCA gains 7.4 and 8.1 points (Table~\ref{tab:res}). Target distillation exceeds both, including when matched to the guided-sample student on 12k training texts or to CCA on its random-class counterfactual.

\textbf{Listening test.} Ten listeners fluent in Mandarin and English compared the distilled student with the baseline and Drama~3, the only external system with all 16 classes, in blind A/B tests on 96 randomly drawn test sentences (three per language and class), with order and sides shuffled per listener. All systems re-synthesized these sentences, cloning one reference voice per language. Listeners saw the tagged text and judged NV accuracy and naturalness, with ties allowed (Table~\ref{tab:listen}). For NV accuracy, the student led the baseline by 32.8 points and Drama~3 by 25.6, with 95\% CIs over sentences and listeners of 20.2 to 44.8 and 9.8 to 42.1 (sign tests over sentences, $p{<}0.001$ and $p{=}0.007$). The preference for the student's NVs supports the improvement measured by the recognizer and judge. Neither comparison showed a significant naturalness preference.

\begin{table}[t]
\centering
\caption{Blind A/B preferences (\%) for NV accuracy and naturalness. Ten listeners judged 96 sentences, giving 960 judgments per system pair for each criterion. Bold marks a significant preference ($p{<}0.05$).}
\label{tab:listen}
\footnotesize
\setlength{\tabcolsep}{3.5pt}
\begin{tabular}{lcccccc}
\toprule
& \multicolumn{3}{c}{NV} & \multicolumn{3}{c}{Naturalness} \\
\cmidrule(lr){2-4}\cmidrule(l){5-7}
Ours vs. & Ours & Tie & Other & Ours & Tie & Other \\
\midrule
Baseline & \textbf{48.1} & 36.6 & 15.3 & 36.5 & 38.6 & 24.9 \\
Drama 3 & \textbf{56.7} & 12.3 & 31.0 & 42.3 & 19.2 & 38.5 \\
\bottomrule
\end{tabular}
\end{table}

\section{Why likelihood training falls short}
\label{sec:tf}
\begin{figure}[t]
\centering
\includegraphics[width=\columnwidth]{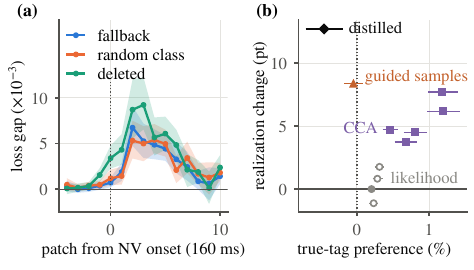}
\caption{Teacher-forced loss with shared noise and flow times (32 draws). (a)~Excess per-patch loss under altered text relative to the true text, aligned at NV onset (baseline evaluated on 400 training utterances, mean with 95\% band, loss $\approx$0.8). (b)~Each point compares the loss increase from replacing the true tag ($x$, mean over fallback and random replacements on 83 held-out recordings, \% of loss, 95\% CI) with the change in test-set realization from baseline ($y$, mean across evaluators). The filled grey point denotes the baseline.}
\label{fig:tf}
\end{figure}

We measured teacher-forced loss on real recordings after changing the tag class, deleting the tag or replacing one word. Each comparison holds the recorded history fixed. Aligning the loss differences to NV onset shows when changing the text affects prediction of the next patch. On 400 training utterances, changing the class raised the baseline's loss by one-sixth to one-quarter as much as replacing a word. Most of the tag effect occurred in the first second after NV onset (Fig.~\ref{fig:tf}a).

On 83 held-out recordings, likelihood training increased the loss gap between true and fallback tags from 0.15\% to 0.29\% of utterance loss within 3,000 steps, with little change in realization (Fig.~\ref{fig:tf}b). CCA increased the random-class gap about eightfold and the fallback gap 2.7-fold, gaining 7 to 8 points in realization. Distilled students achieved the largest gains despite having lower loss under the fallback tag than under the true tag. Across these models, a stronger teacher-forced preference for the true tag did not reliably predict the class produced at synthesis.

The timing in Fig.~\ref{fig:tf}a suggests an explanation. Before onset, the text can distinguish a requested laugh from a sigh even when the preceding speech fits either. Once the first recorded patches of a laugh enter the history, both text conditions are evaluated on a history that already sounds like laughter. Continuing that recording can therefore incur a similar loss under both tags. During synthesis, an initial sigh would instead provide acoustic evidence for continuing a sigh. A loss computed along the recorded trajectory does not measure how often the model enters either trajectory from the same prefix.

Likelihood fine-tuning still fits the recorded patches, including when they are reweighted. Guided samples~\cite{seqkd} and CCA~\cite{cca} shift training toward a guided distribution. Target distillation incorporates the teacher's response to the tag substitution directly into the velocity the student learns (Eq.~\ref{eq:distill}). Figure~\ref{fig:tf}b scores every student against the original data velocity. A student trained toward a shifted target can therefore improve the sampled class without widening the loss gap on the unmodified recording.

\section{Discussion and conclusion}
In VoxCPM2, a tag largely determines where a vocalization occurs; most errors concern which one. Section~\ref{sec:tf} suggests why likelihood fine-tuning helped little: once a recorded NV begins, its audio already reveals the class, so the loss no longer tests which class the model would choose. What helped was not stronger CFG but holding the words fixed and changing only the tag. Distilled into the training target, this contrast raised realization under both evaluators with plain decoding, at a small lexical cost. Listeners preferred the student's NVs in both comparisons, with no significant naturalness difference.

Gains vary by class. For [breath] and [sigh], the fallback classes excluded from Table~\ref{tab:diag}, the recognizer and listeners (not the judge) favored the baseline, and [gasp] fell under both evaluators. Ten bilingual listeners compared the student only with the baseline and Drama~3, using cloned voices; other systems and NV-Bench were scored automatically.


\end{document}